\documentclass{elsart}
\usepackage{graphics}
\usepackage{graphicx}
\usepackage{epsfig}
\usepackage{amssymb}

\begin{document}
\begin{frontmatter}
\title{A dynamic hybrid model based on wavelet and fuzzy regression
for time series estimation}
\author{Olfa Zaafrane}
\ead{zaafraneolfa@yahoo.fr}
\address{Department of Quantitative Methods, Faculty of Economic Sciences
and Management, Sidi Messaoud, 5111 Hiboun, Mahdia, Tunisia.}
\author{Anouar Ben Mabrouk\thanksref{label1}}
\thanks[label1]{Corresponding Author}
\ead{anouar.benmabrouk@issatso.rnu.tn}
\address{Computational Mathematics Laboratory, Department of
Mathematics, Faculty of Sciences, 5019 Monastir, Tunisia.}
\begin{abstract}
In the present paper, a fuzzy logic based method is combined with
wavelet decomposition to develop a step-by-step dynamic hybrid model
for the estimation of financial time series. Empirical tests on
fuzzy regression, wavelet decomposition as well as the new hybrid
model are conducted on the well known $SP500$ index financial time
series. The empirical tests show an efficiency of the hybrid model.
\end{abstract}
\end{frontmatter}
\section{Introduction}
The study of time series is an interesting task especially in
financial contexts such as modeling, estimating, approximating and
prediction. It necessitates a precise and deep comprehension of the
series characteristics for a suitable choice of the model to be
applied. The estimation process guarantees the detection of passed
disfunction causes and therefore, it helps to take the eventual and
possible precautions at the suitable time. A fine and preventive
analysis guarantees a good preparation for the future and a robust
prediction in front of random breaks and non anticipated changes.
Financial time series are for example, are characterized by very
specific stylized facts where a respect with estimation method
proves its efficiency. Observing the distribution tail, for the
leptokurtic cases always evaluated by the kurtosis, the series
values far from the mean of the series appears with probabilities
that overcome the normal distribution. In financial case, the
studies have shown that the tail distribution is not leptokurtic but
in the contrary, it has a kurtosis exceeds the normal case.
Furthermore, observing the volatility clustering, financial time
series are characterized by complex combinations of components with
high frequencies. These facts are somehow due to the presence of the
random or stochastic behavior of the markets. Besides, the market
may be characterized by infinite volatility allowing long memory
process. This induces the appearing of scaling law invariance on the
volatility (Walter, 2001). Indeed, Walter expects that the
conciliation between absence of long memory on profitability and its
presence on volatility is a modeling financial problem. Due to these
facts, some classical methods have been classified as incapable to
analyze financial series. ARCH and GARCH models did not take into
account the kurtosis degree of the series. Furthermore, ARCH model
and its terminologies have attained their limits in the field of
financial modeling due to the fact that the scaling law in
volatility has not been included in the model. (See also Walter,
2001). For this aim, researchers in financial time series field have
thought to introduce other methods that may induce more efficient
models and to understand some aspects of non stationary,
auto-regression, filtering, support vector machine models and
prediction, neural networks models and predicting. See (Angue,
2007), (Azizieh, 2002), (Ben Mabrouk et al 2008a,b), (Ben Mabrouk et
al 2008), (Ben Mabrouk et al 2010), (Ben Mabrouk et al, 2011),
(Chang et al, 2001), (Chen et al, 2006), (Chou, 2005), (Klir et al,
1995), (He et al, 2007), (Khashei et al, 2008), (Kim et al, 1996),
(Mitra et al, 2004), (Podobnik et al, 2004), (Ramsey, 1999),
(Struzik, 2000), (Tanaka et al, 1982), (Tseng et al, 1999), (Tseng
et al, 2001), (Wang et al, 2000), (Watada, 1992), (Wu et al,2002),
(Zopoundis et al, 2001).

In the present paper, one aim is to apply wavelet theory and fuzzy
logic theory to develop an estimation model for financial series. We
search to judge the efficiency of fuzzy regression to estimate
financial series. Next, we apply the discrete wavelet decomposition
which improve especially the study of the local behavior of the
series. Comparing the two methods of estimation, we have discovered
that an hybrid model combining wavelet estimation with fuzzy logic
estimation is possible. We then developed such a model which takes
into account the non stationary behavior of the series as well as
its local fluctuations and its fuzzy characteristics. The model
combines wavelet decomposition with fuzzy regression. Next, an
empirical study based on the famous $SP500$ index is provided in
order to improve the theoretical parts.

The present paper is organized as follows. A first section is
devoted to the presentation of the series characteristics. Section
2 is devoted to the development of the fuzzy regression model for
the estimation of financial time series. In section 3, a wavelet
analysis of time series is provided. In section 4, the hybrid
model deduced by combining fuzzy logic with wavelet decomposition
is developed. Finally, an empirical study on the SP500 index is
developed in section 5 leading to a comparison between the
different models and improving the impact of the hybrid scheme.
\section{The Data Description}
In the present paper, we propose to study the behavior of the well
known financial index $SP500$ which is a stock index describing the
fluctuations of the stock capitalization due to the 500 most large
economic societies of the American stock. It is composed of a number
of 380 industrial firms, 73 financial societies, 37 public service
firms, and 10 transport ones. The choice of such an index is
motivated essentially by its central role as a measure of the
American economy performance. Besides, the international financial
integration is often increasing which forces the international
exchanged productions to be strongly related. So that, as the
American market is the center of international transactions, any
variation of its index such as $SP500$ immediately affects on other
external markets. Furthermore, the study of the USA market index is
of interest nowadays due to the financial international crisis which
has been started from this market and next affected the world-wise
markets. So, searching a good solution to understand the crisis is
of priority.

The data basis consists of $SP500$ index monthly values during the
period from August 1998 to March 2009 allowing a basis of size
$N=128=2^7$. We applied the $\log$-values of the series in order to
reduce the range of the series. The statistic characteristics of the
series are resumed in the following table.
\begin{table}[ht]
\centering
\begin{tabular}{|l|l|}
\hline\hline
Sample size, $N=$&128\\
\hline
Mean&7.0921\\
\hline
Variance&0.0246\\
\hline
Maximum&7.3456\\
\hline
Minimum&6.6000\\
\hline
Kurtosis&3.2229\\
\hline
skewness&-0.9017\\
\hline\hline
\end{tabular}
\caption{Statistic Characteristics}
\end{table}

We notice a kurtosis value over-crossing the normal
value 3 which means that the series is leptokurtic. The skewness of
the series induces a negative value which means that the data are
spread out more to the left relatively to the means of the series
than to the right. The following figure represents the original
series $S(t)=\log(SP(t))$, where $SP(t)$ is the corresponding value
of the index $SP500$ at the month $t$.
\begin{figure}[htbp]
\begin{center}
\epsfig{figure=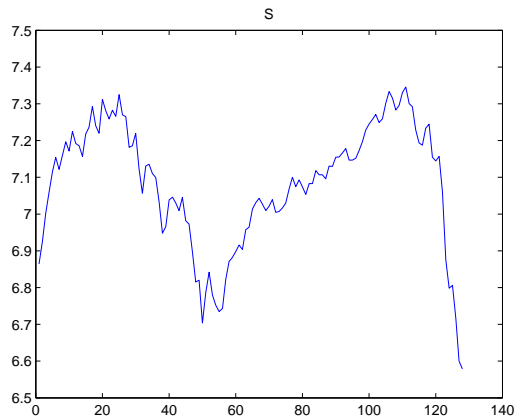,width=8cm}\caption{Original Series
$S(t)$}\label{figuresignalreel}
\end{center}
\end{figure}
\section{A fuzzy regression model}
The reasons behind the test of fuzzy regression for modeling
financial series has many justifications. Firstly, financial series
have always an  ambiguous relation concerning dependent variables
and independent one; The time variable here. Such an ambiguity is
not taken into account in almost all statistical methods, but in the
contrary they assume that the behavior is always definite.
Furthermore, financial series such as $SP500$ are already fluctuated
with an unpredicted behavior. This permanence makes the future
values of the series to be fuzzy and/or imprecise. The fuzzy
regression was already applied as a privileged method for the
estimation of uncertain and imprecise data. See (He et al, 2007),
(Khashei et al, 2008), (Kim et al, 1996), (Sanchez et al, 2003),
(Shapiro, 2005), (Terence, 1999), (Tseng et al, 1999), (Tseng et al,
2001), (Watada, 1992), (Wu et al, 2002), (Zopoundis et al, 2001).

\begin{figure}[htbp]
\begin{center}
\epsfig{figure=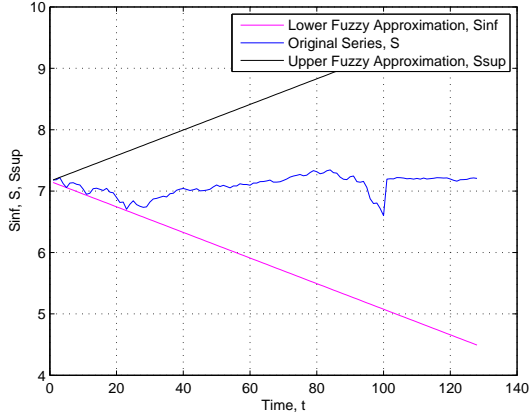,width=8cm} \caption{Original series
and its fuzzy regression estimation}\label{figureregressionfloue}
\end{center}
\end{figure}

In this section, a fuzzy regression model is applied to estimation
the $SP500$ index series. The model due to Watada 1992, is applied
here. This model is reviewed hereafter. It is based on the following
fuzzy linear programming.
\begin{equation}
\left\{\matrix{
\medskip\,MinS=s_0+s_{1}\hfill\cr\medskip
s.t\hfill\cr\medskip c_{0}+c_{1}t_{i}-(1-h)(s_{0}+s_{1}|t_{i}| )
\leq Y_{i},\hfill\cr\medskip
c_{0}+c_{1}t_{i}+(1-h)(s_{0}+s_{1}|t_{i}|) \geq
Y_{i},\hfill\cr\medskip s_{j}\geq0 \quad and \quad
s_{1}\geq0,\hfill\cr\medskip \forall t_{i}=1,...,128.\hfill}\right.
\end{equation}
where
\begin{itemize}
\item[-] $h$ is a standard threshold, hereafter applied for $h=0.5$.
\item[-] $a_{j}=(c_{j}, s_{j})$, $(j=0,1)$ is a triangular fuzzy number.
\item[-] $t_{i}$ is the time variable.
\item[-] $Y_{i}$ is the observed index value at the time $t_i$, $i=1,\dot,128$.
\end{itemize}
The problem is resolved using the Software LINGO9 resulting in the
following fuzzy coefficients (Triangular fuzzy numbers).
\begin{equation}
a_{0}=(6.887995,0)\quad\hbox{and}\quad\,a_{1}=(0.01066521,0.06912872).
\end{equation}
As a result the lower and upper estimations of the index series is
provides resulting in the following fuzzy regression equation.
\begin{equation}
Y_{i}=(6.887995, 0)+ (0.01066521, 0.06912872)*0,5*t_{i}.
\end{equation}
The original series with its fuzzy estimation are shown in the
Figure \ref{figureregressionfloue} following.

We notice that although the fuzzy regression model takes into
account the uncertain behavior of the information, it did not fits
well the tendency of the series, and it assumes that a monotone
behavior exists which means that it ignores the fluctuations already
characterizing the data. Besides, the error estimation is important
resulting in the values
$$
MSE=5.31016565\quad\hbox{and}\quad RMSE=2.30437967
$$
where
\begin{equation}
MSE=\sum_{i}{(Y_{i}-\widehat{Y_{i}})^{2}/128}\quad i=1,...,128,
\end{equation}
\begin{equation}
RMSE=\sqrt{MSE},
\end{equation}
and $\widehat{Y_{i}}$ is the estimated value of the index at the
time $t_i;\,i=1,...,128$.

As a conclusion, the fuzzy regression has been proved to be
incapable for a robust estimation with a least error for the series
applied. It necessitates to be corrected to fit the fluctuations and
then the random behavior of the series. So, an analysis permitting
to localize these fluctuations is necessary. It consists of wavelet
analysis which will be developed in the next section.
\section{Wavelet analysis of the series}
Wavelet analysis is always applied to show how the series is
volatile, and then to detect eventual fluctuations, (Patick, 2005).
Wavelet analysis permits also to represent the strongly fluctuated
series without necessitating a knowledge of the explicit functional
dependence.  Such a capacity is of great role especially for
financial time series where such a dependence is always unknown.

We propose hereafter to conduct a wavelet analysis of the series due
to the index $SP500$ in order to localize well the fluctuations of
the series. A maximum level decomposition $J=6$ is fixed allowing a
decomposition or a projection on the approximation space $V_6$
relatively to a Daubechies $DB4$ multi-resolution analysis with
Matlab7 software.

As a result the series $S(t)$ is decomposed on the form
$$
S=(A_6,\quad D_1,\quad D_2,\quad D_3,\quad D_4,\quad D_5,\quad D_6)
$$
or equivalently,
$$
S=D_1+D_2+D_3+D_4+D_5+D_6+A6
$$
where $A_6$ is the global form of $S(t)$ at the level $6$ called
also the trend or tendency, and $D_1$, $D_2$, $D_3$, $D_4$, $D_5$
and $D_6$ are the detail components of $S(t)$ obtained by projecting
the series on the detail spaces $W_1$, $W_2$, $W_3$, $W_4$, $W_5$
and $W_6$. These components are represented hereafter.
\begin{figure}[http]
\begin{center}
\epsfig{figure=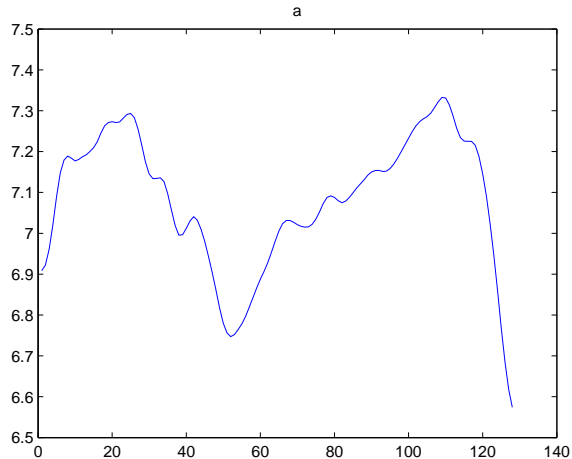,scale=0.6}\caption{Approximation
$A_6$}\label{figurea}
\end{center}
\end{figure}
\begin{figure}[http]
\begin{center}
\epsfig{figure=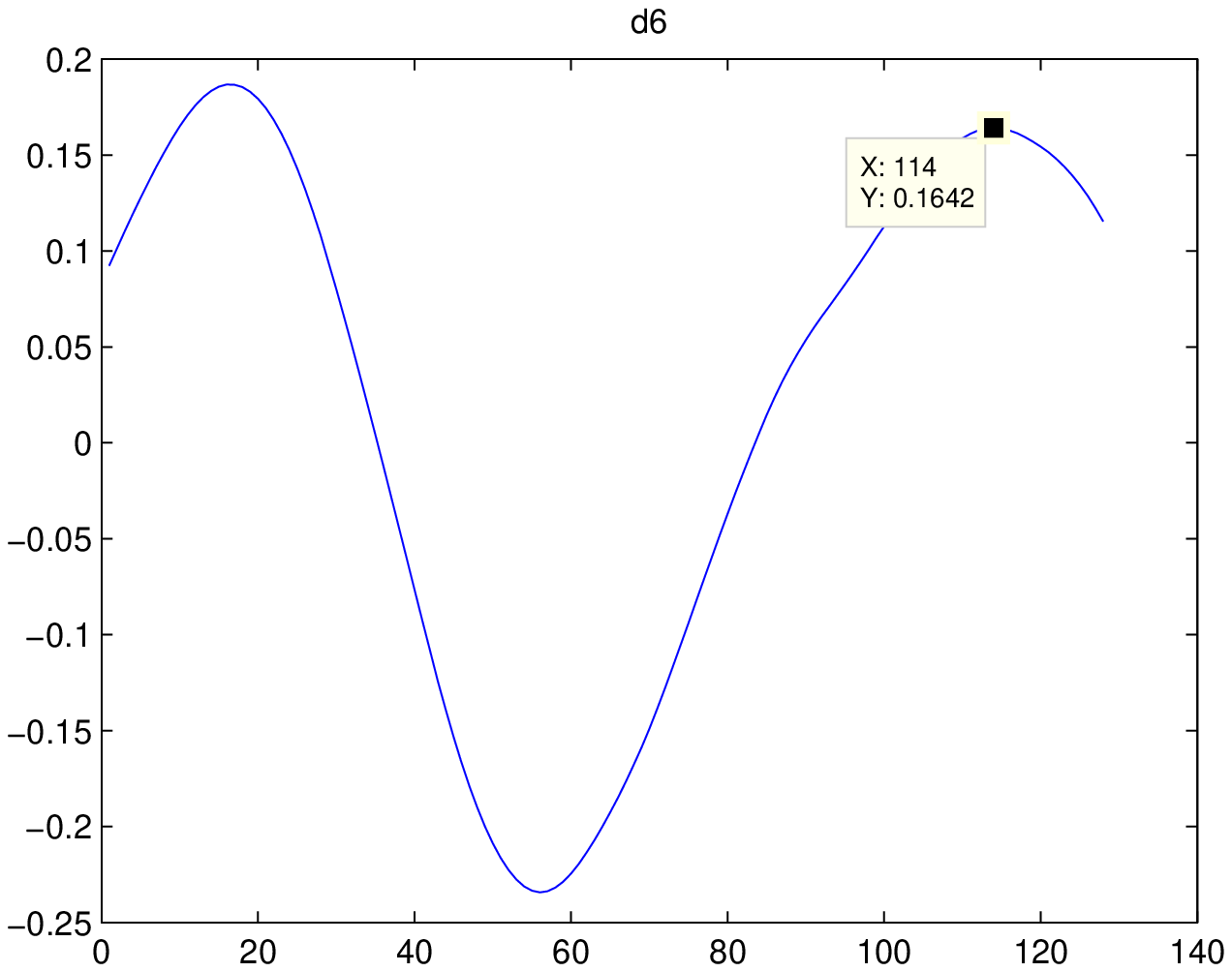,scale=0.6}\caption{Detail component
$D_6$}\label{figureD6}
\end{center}
\end{figure}
\begin{figure}[http]
\begin{center}
\epsfig{figure=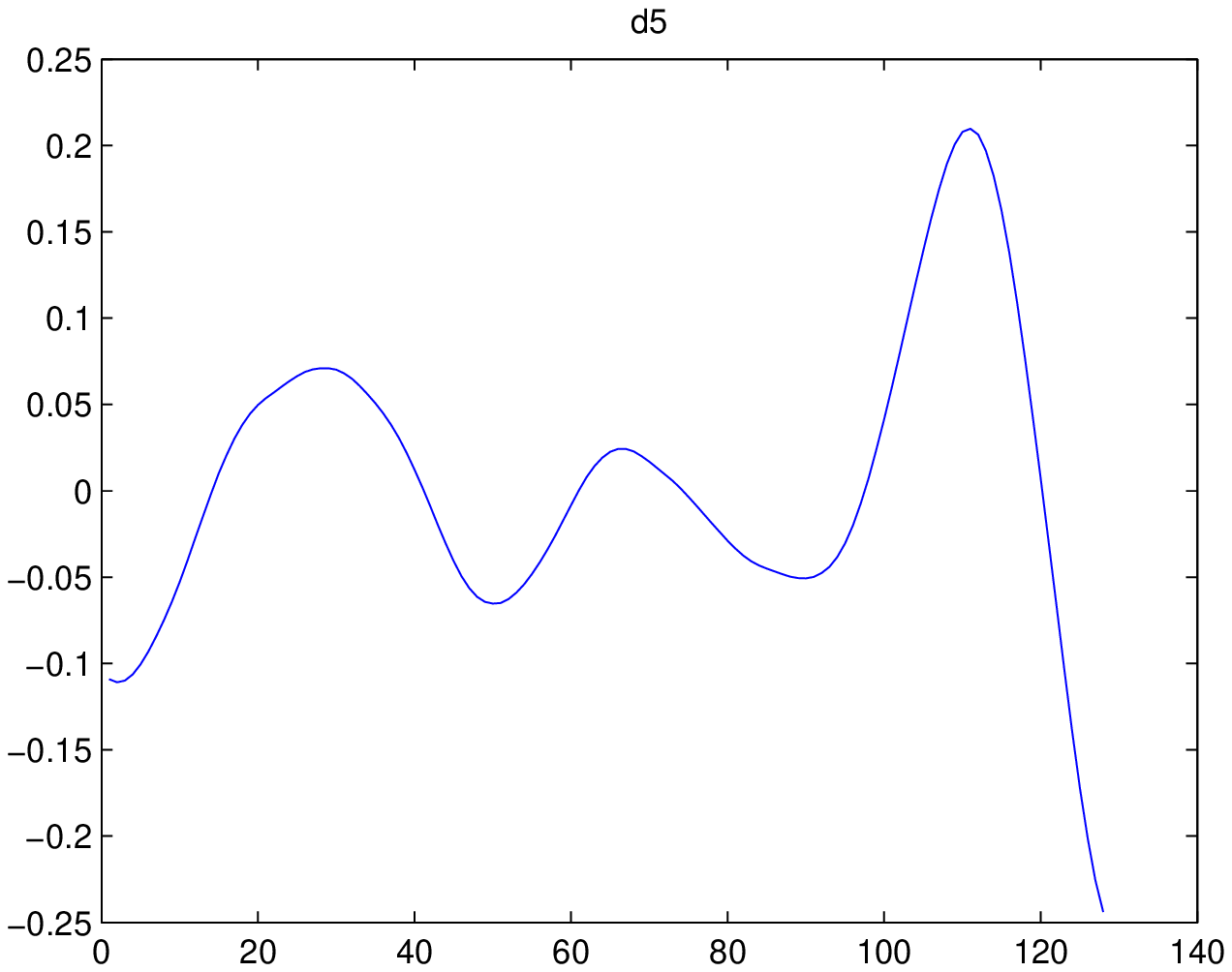,scale=0.6}\caption{Detail component
$D_5$}\label{figureD5}
\end{center}
\end{figure}
\begin{figure}[http]
\begin{center}
\epsfig{figure=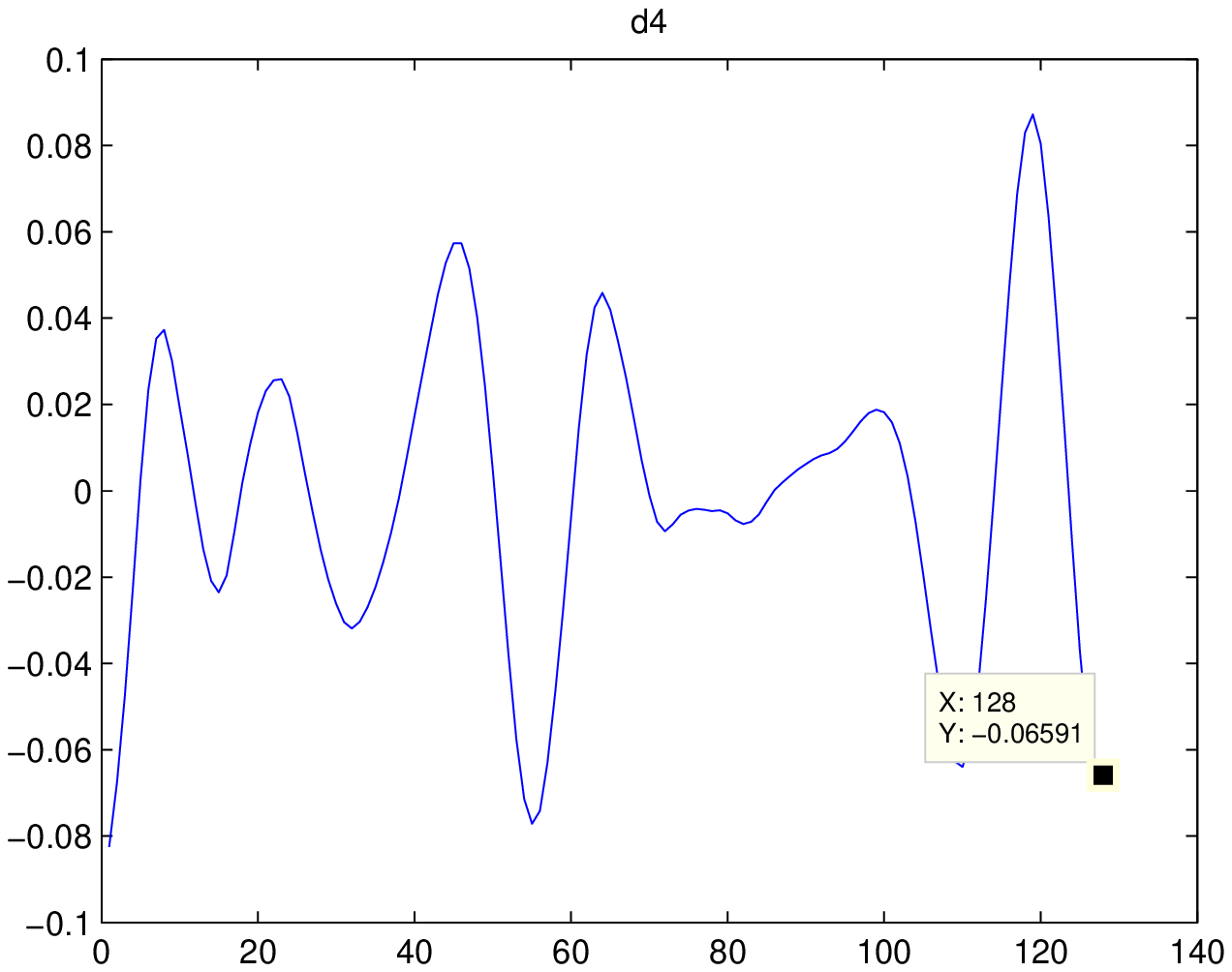,scale=0.6}\caption{Detail component
$D_4$}\label{figureD4}
\end{center}
\end{figure}
\begin{figure}[http]
\begin{center}
\epsfig{figure=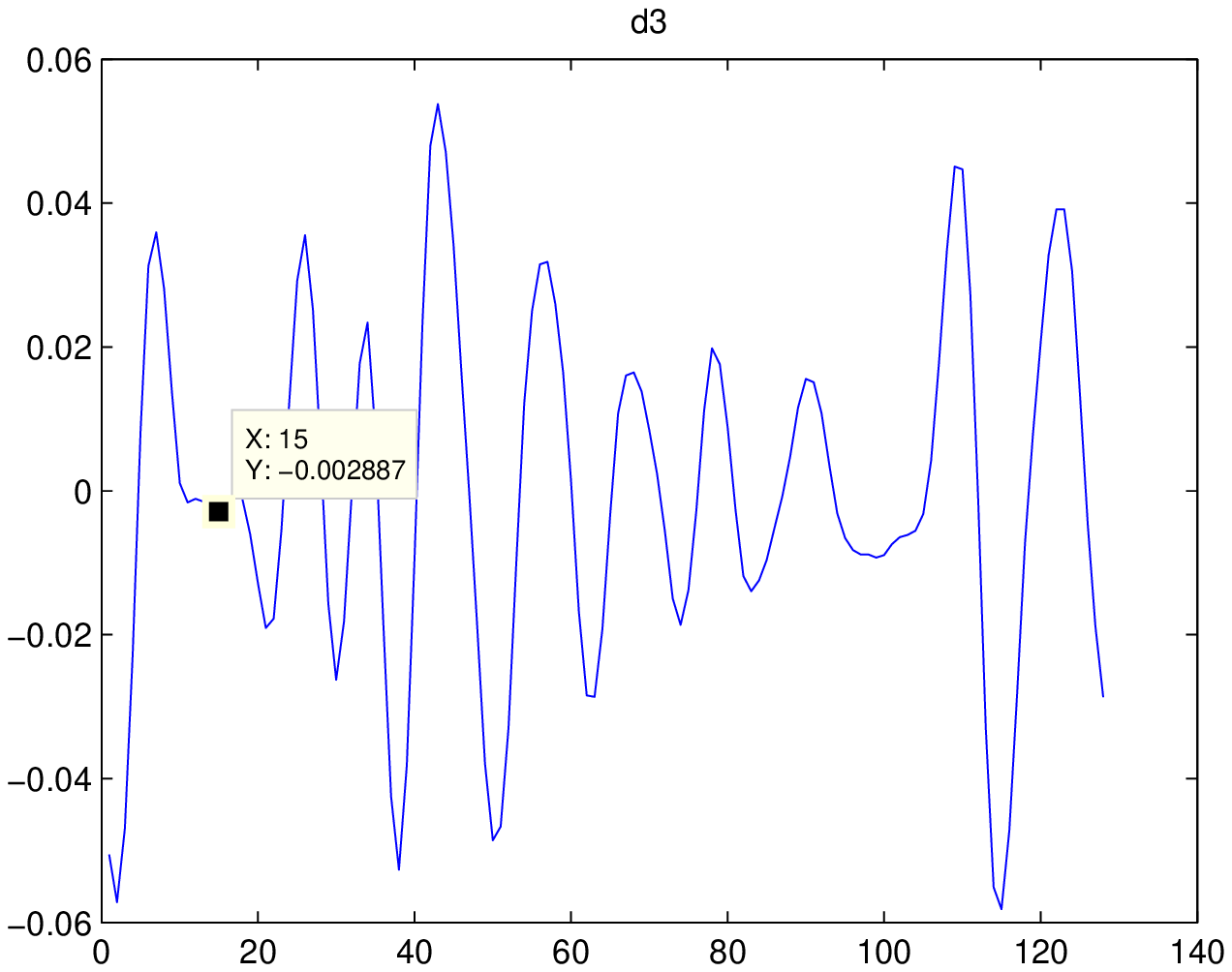,scale=0.6}\caption{Detail component
$D_3$}\label{figureD3}
\end{center}
\end{figure}
\begin{figure}[http]
\begin{center}
\epsfig{figure=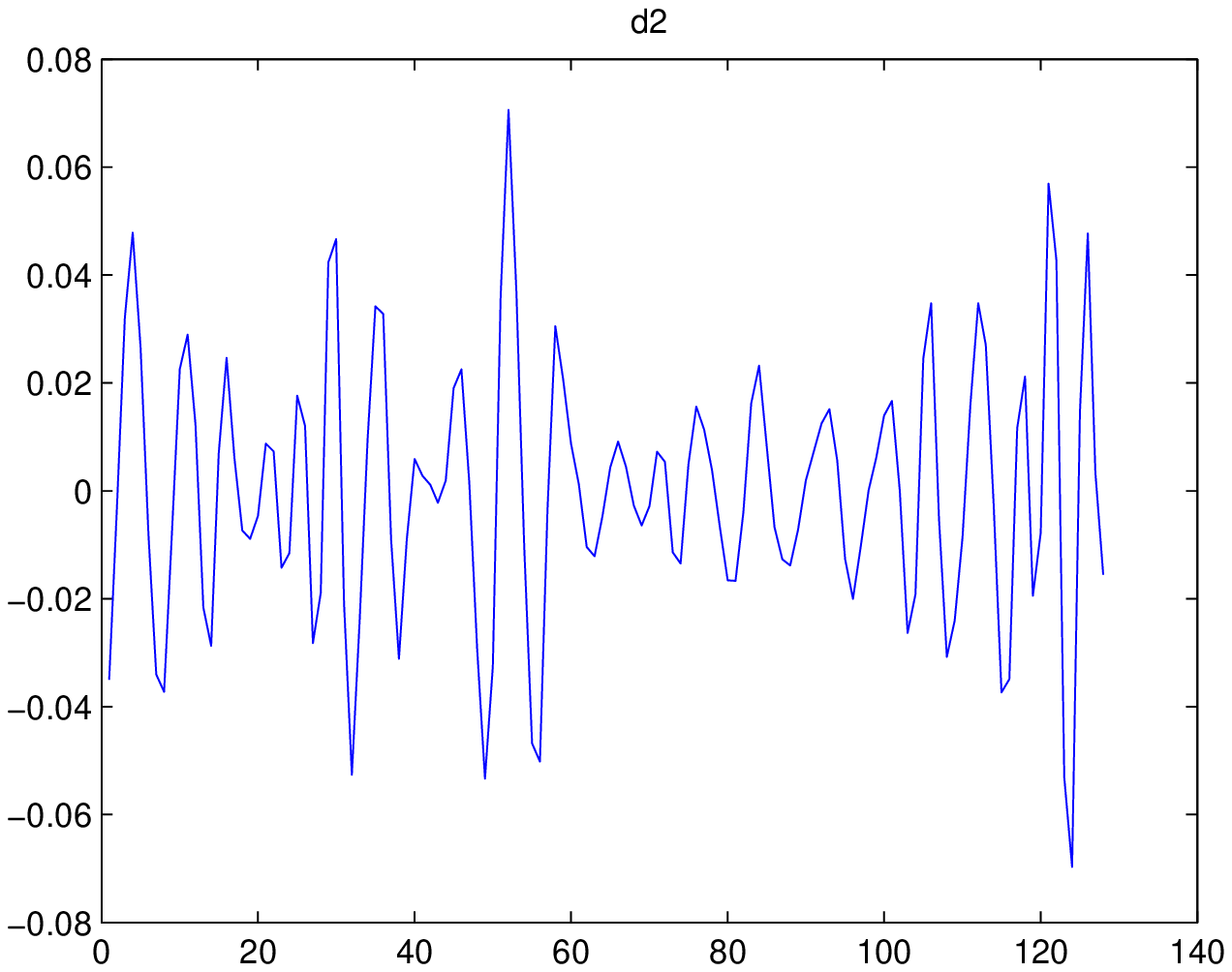,scale=0.6}\caption{Detail component
$D_2$}\label{figureD2}
\end{center}
\end{figure}
\begin{figure}[http]
\begin{center}
\epsfig{figure=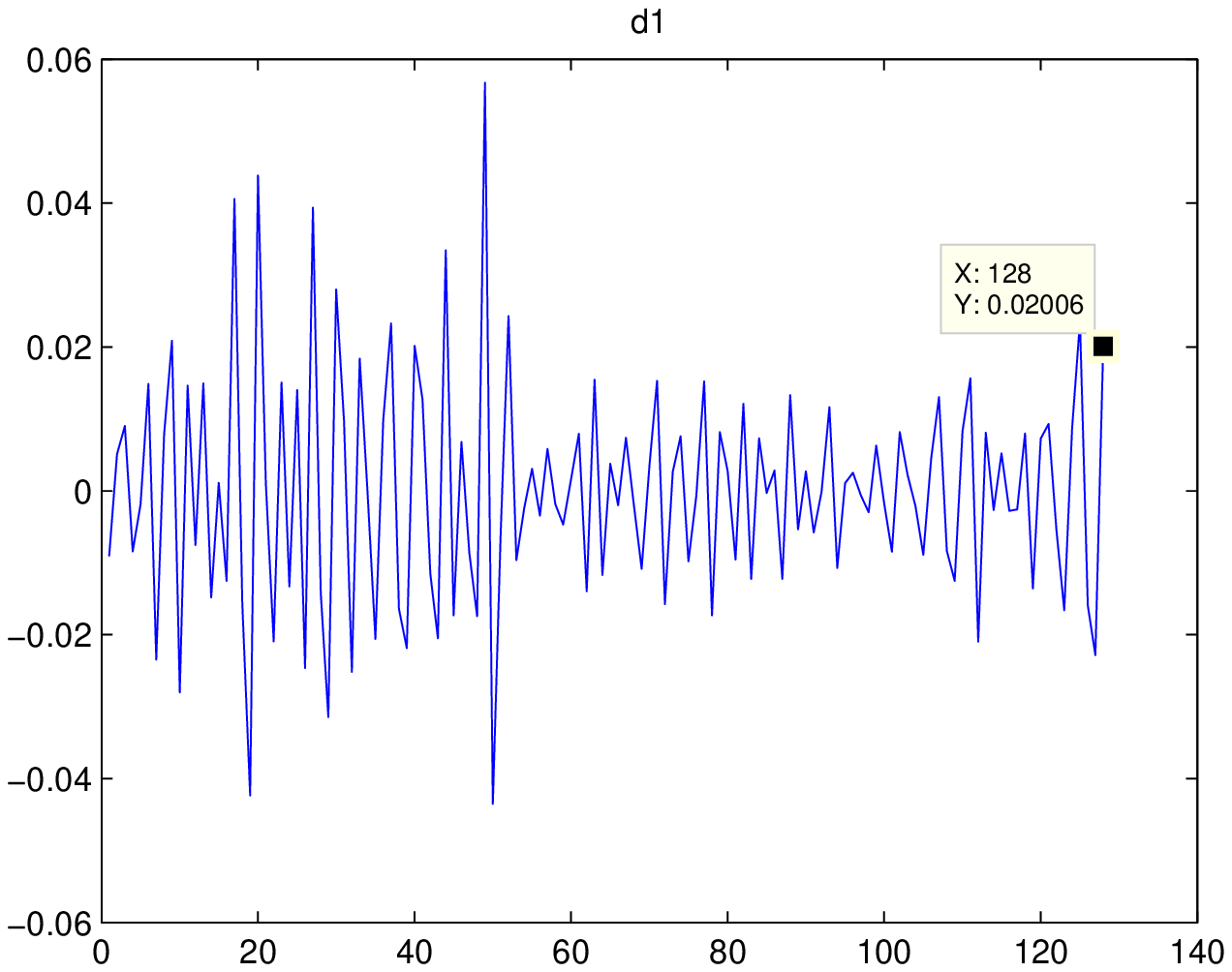,scale=0.6}\caption{Detail component
$D_1$}\label{figureD1}
\end{center}
\end{figure}

We notice easily from these figures the localizations of the
fluctuations of the series. The component $A_6$ shows the low
frequency fluctuations. The components $D_i$, $i=1,2,...,6$
represents the high frequency behavior. We remark that the series is
more fluctuated at detail levels $D_4$ and $D_3$ more than $D_5$ and
$D_6$. The volatile aspect of the series is clearly observed from
$D_1$ and $D_2$.
\section{Hybrid estimation model}
As we have localized the fluctuations of the series, we propose to
return to the fuzzy regression model and to conduct a correction on
it consisting in re-developing a dynamic fuzzy regression taking
into account both the fluctuations and the uncertain aspect of the
series. Denote $S(t)$ the financial time series due to the $SP500$
index introduced previously. The proposed hybrid model is described
by the following steps.
\begin{itemize}
\item {\bf Step 1:} The wavelet decomposition of the series; ($D_{1}$,$D_{2}$,$D_{3}$,$D_{4}$,$D_{5}$,$D_{6}$,$A_6$).
\item {\bf Step 2:} Compute the localizations of the extremum points of
each component $D_i$; $i=1,2,...,6$.
\item {\bf Step 3:} Apply the fuzzy regression to estimate the
restriction of the series on each interval $[t_n(i),t_{n+1}(i)]$
where $t_n(i)$, $t_{n+1}(i)$ are two consecutive extremum points for
the component $D_i$; $i=1,2,...,6$.
\item {\bf Step 4:} For all $i=1,2,...,6$, regroup the new series obtained on the whole time interval $\displaystyle\bigcup_{n}[t_n(i),t_{n+1}(i)]$.
\end{itemize}
We remark easily that the proposed model fits the peace-wise
monotonicity of the time series. On each interval, where the series
is monotone the fuzzy regression is applied with corresponding fuzzy
numbers. The results due to this model are shown in following
figures.
\begin{figure}[htbp]
\begin{center}
\epsfig{figure=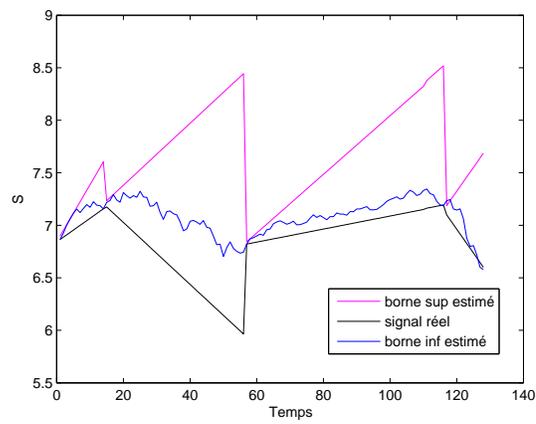,width=8cm}\caption{Estimation
relatively to $D_6$.}\label{figuredetailestimation6}
\end{center}
\end{figure}
\begin{figure}[htbp]
\begin{center}
\epsfig{figure=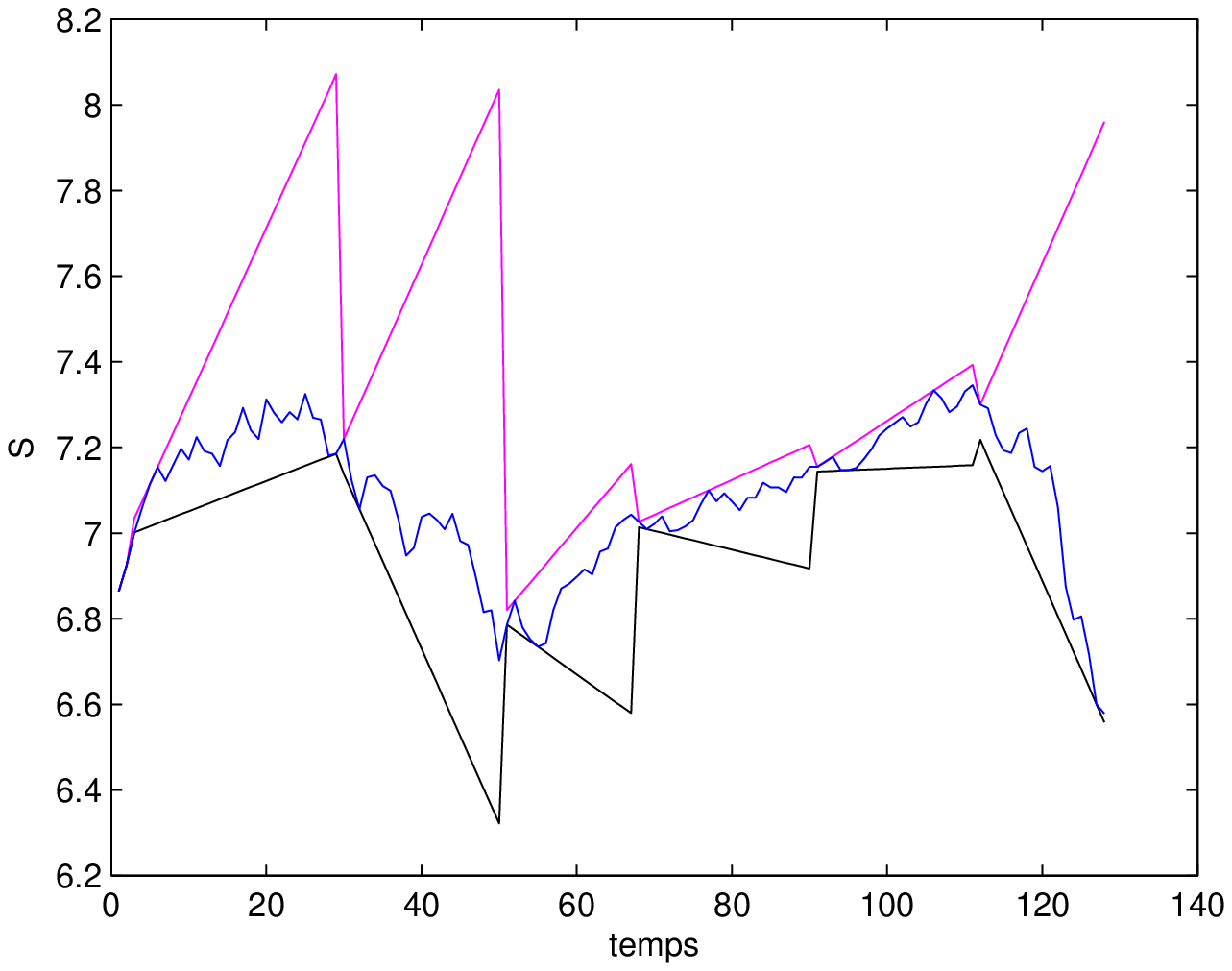,width=8cm}\caption{Estimation
relatively to $D_5$.}\label{figuredetailestimation5}
\end{center}
\end{figure}
\begin{figure}[htbp]
\begin{center}
\epsfig{figure=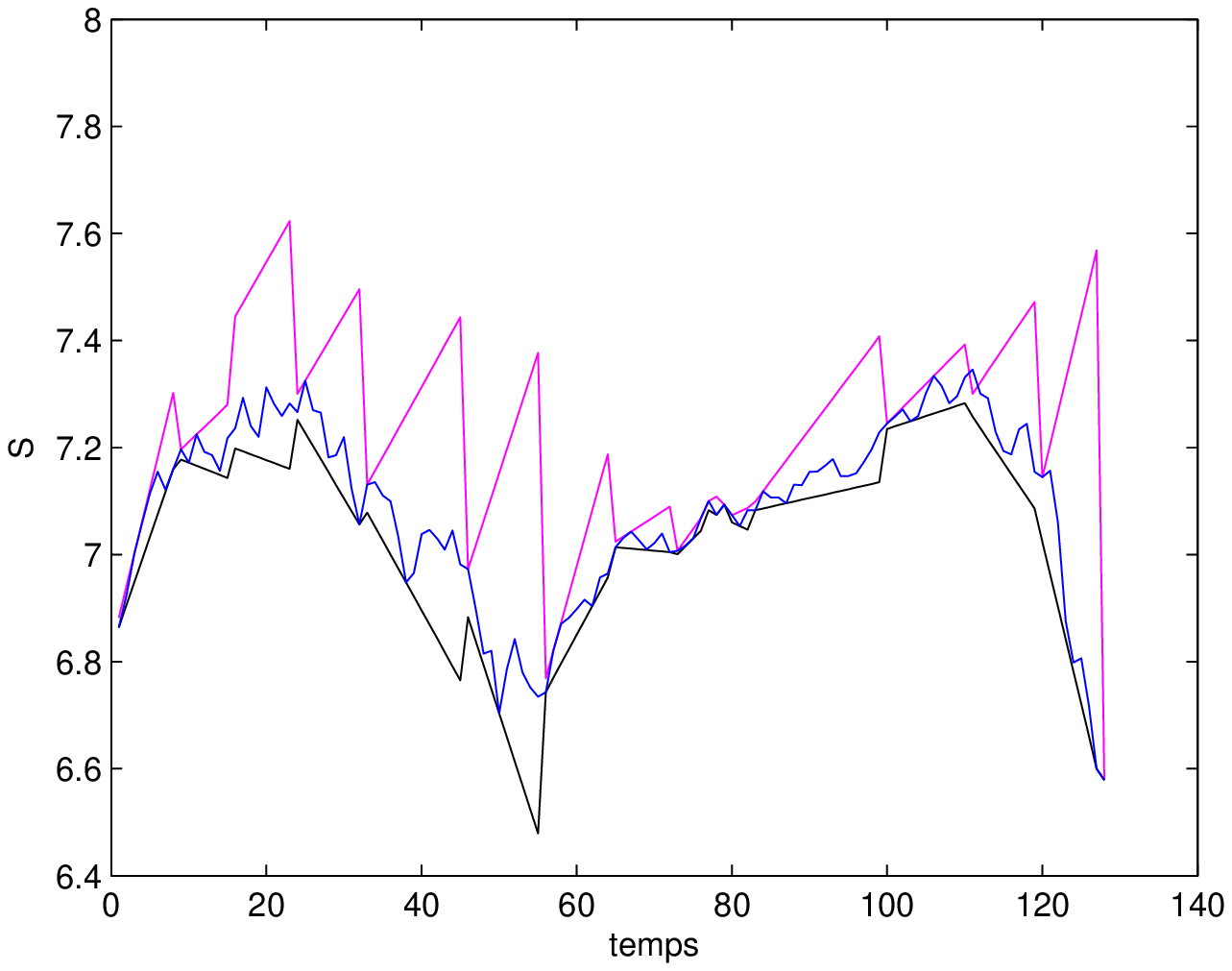,width=8cm}\caption{Estimation
relatively to $D_4$.}\label{figuredetailestimation4}
\end{center}
\end{figure}
\begin{figure}[htbp]
\begin{center}
\epsfig{figure=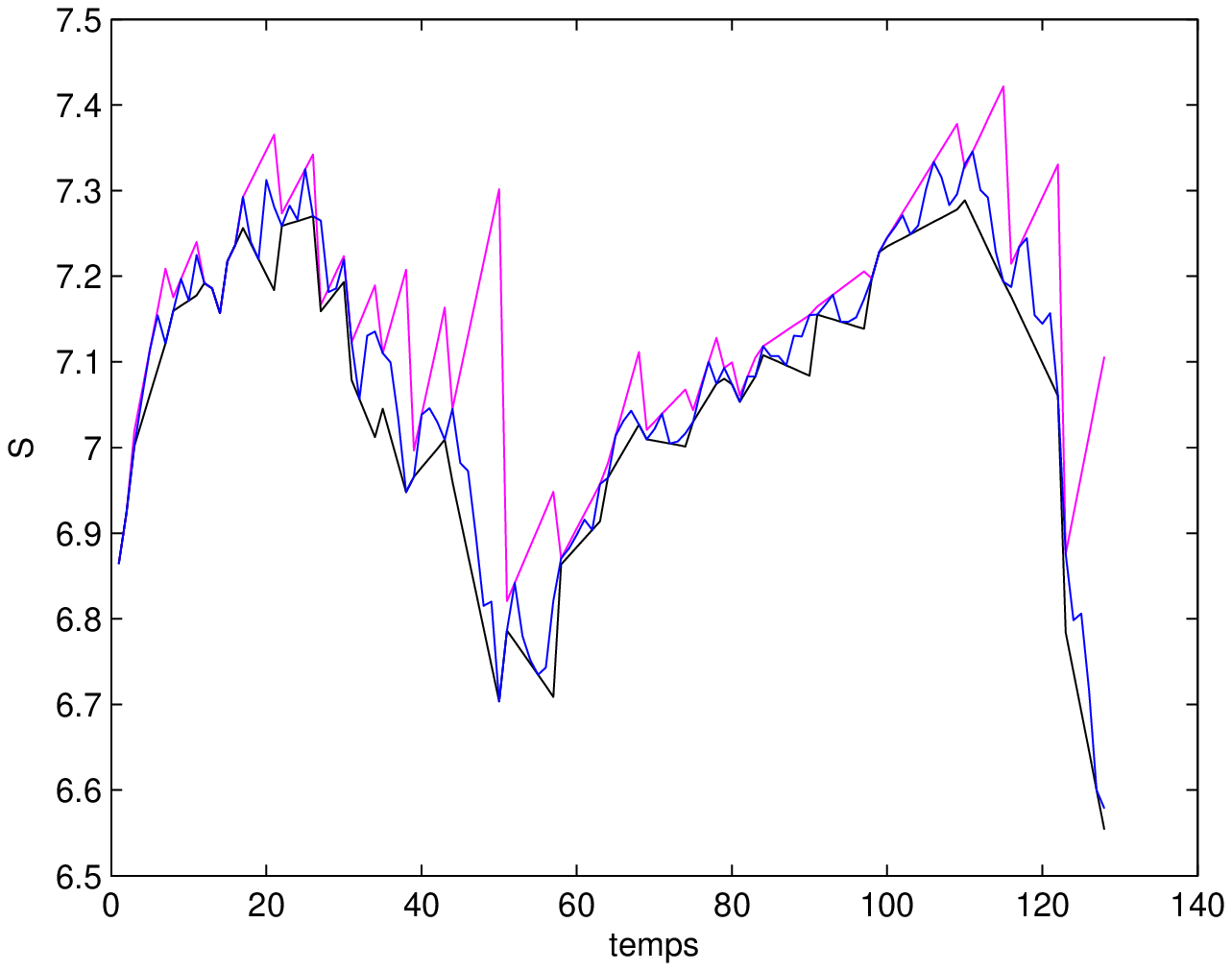,width=8cm}\caption{Estimation
relatively to $D_3$.}\label{figuredetailestimation3}
\end{center}
\end{figure}
\begin{figure}[htbp]
\begin{center}
\epsfig{figure=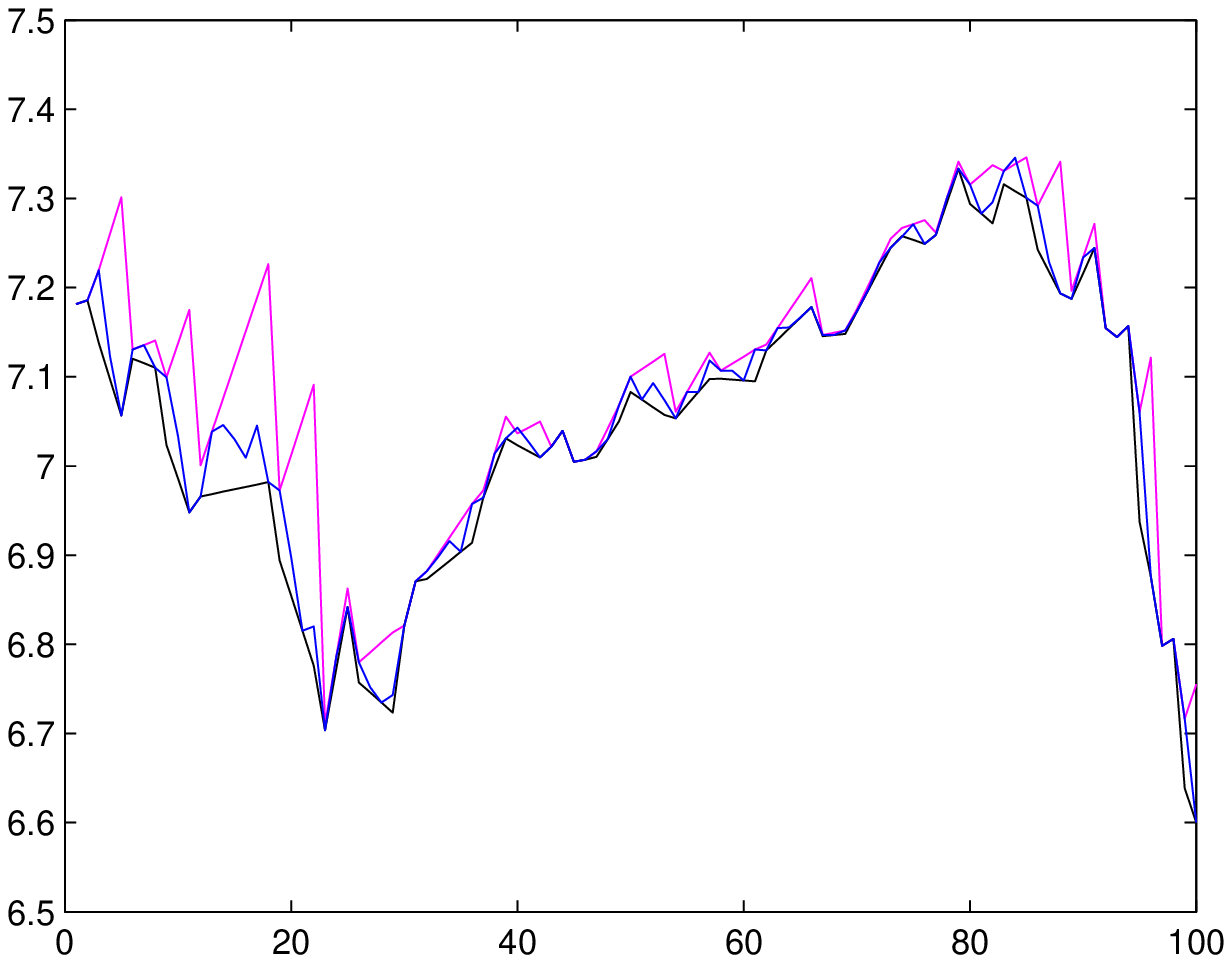,width=8cm}\caption{Estimation
relatively to $D_2$.}\label{figuredetailestimation2}
\end{center}
\end{figure}
\begin{figure}[htbp]
\begin{center}
\epsfig{figure=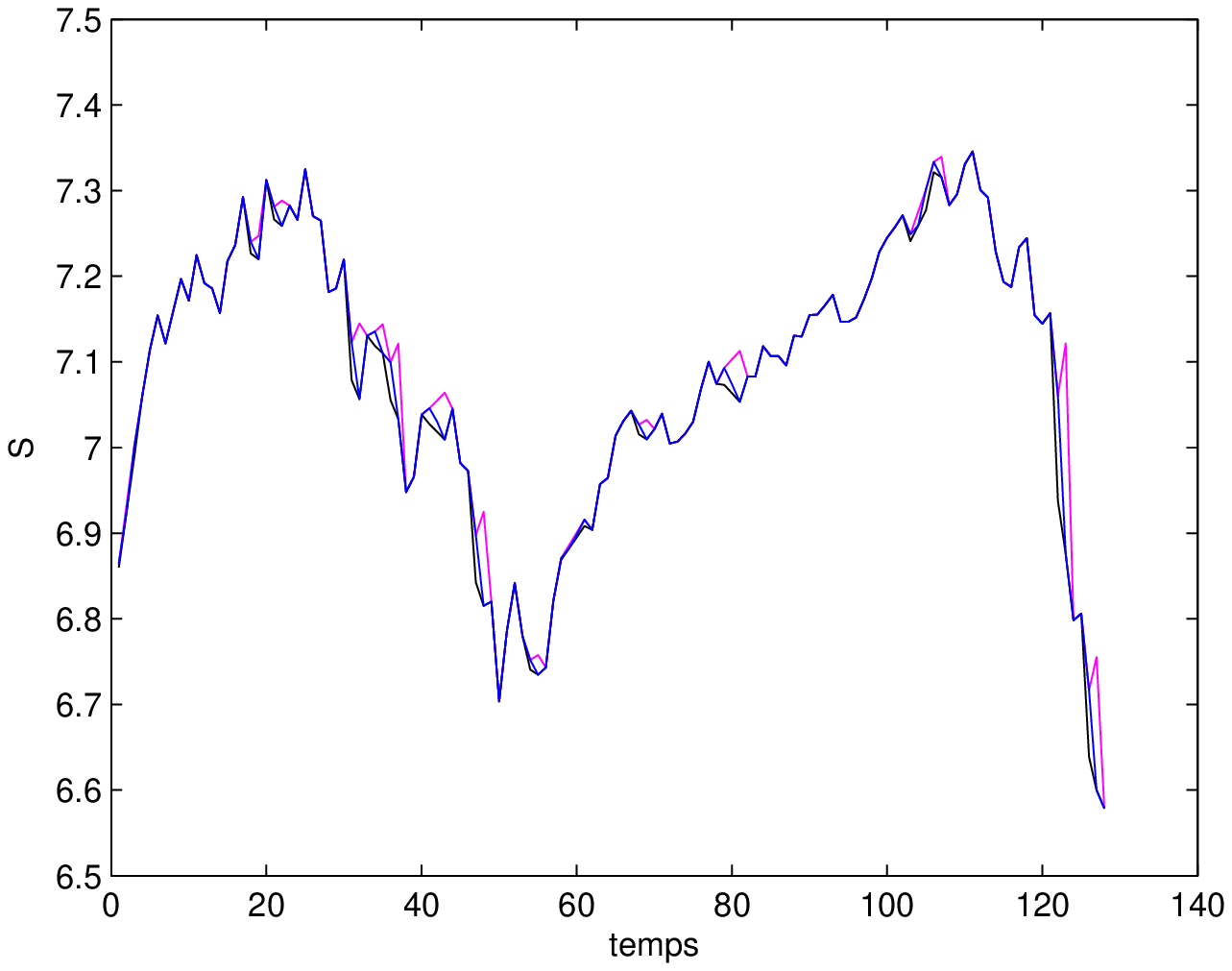,width=8cm}\caption{Estimation
relatively to $D_1$.}\label{figuredetailestimation1}
\end{center}
\end{figure}

As we see, the new estimation due to the hybrid model fits more the
original series as the detail level decreases. Here, we stress the
fact that Daubechies wavelets in the software Matlab7 uses the
frequency index $-j$ contrarily to the theoretical way of wavelet
basis definition which uses instead the index $j$. So, we seek here
an increasing in the detail approximation as $j$ decreases. Indeed,
the estimation relatively to $D_6$ is somehow abusive (See Figure
\ref{figuredetailestimation6}). This is due to the fact this
component does not contain an important number of extremum points or
fluctuations. The estimation becomes more efficient when using $D_5$
(See Figure \ref{figuredetailestimation6}). Next, Figures
\ref{figuredetailestimation4}, \ref{figuredetailestimation3},
\ref{figuredetailestimation2}, \ref{figuredetailestimation1} show an
increasing in the fitness between the original series and the hybrid
model estimated one. This is due to the fact that the hybrid model
follows well the fluctuations of the series. To finish with this
model, we provided in the following table the different error
estimates corresponding to the details $D_i$; $i=1,2,...,6$.
\begin{table}[http]
\centering
\begin{tabular}{|l|l|l|}
\hline\hline
The model&MSE&RMSE\\
\hline
Fuzzy Regression&1.5380&1.2401\\
\hline
Hybrid with $D_6$&0.4100061&0.64032\\
\hline
Hybrid with $D_5$&0.1506692&0.380324\\
\hline
Hybrid with $D_4$&0.0402139&0.2006\\
\hline
Hybrid with $D_3$&0.01133175&0.1064507\\
\hline
Hybrid with $D_2$&0.00261067&0.0611\\
\hline
Hybrid with $D_1$&0.00060889&0.024675\\
\hline\hline
\end{tabular}
\caption{Error estimates}\label{erreur}
\end{table}
\section{Conclusion}
In the present paper, a fuzzy regression estimation is applied to
estimate financial time series. Such estimation is shown to be not
efficient. It gives an estimation with affine boundaries to the
series which did not follow the fluctuations well. As financial time
series are very volatile, a wavelet decomposition is applied next to
localize the fluctuations and then to prepare to a more
sophisticated fuzzy model taking into account the fluctuations. As a
result, an hybrid model combining fuzzy regression and wavelet
decomposition is developed. Finally, the different models are tested
on the well known financial time series of the $SP500$ index. The
empirical tests show an efficiency of the hybrid model. We intend in
the future to apply the hybrid method or modified versions for other
time series and for prediction aims.

\end{document}